\documentclass[aps,preprint,a4paper,floatfix,superscriptaddress,showpacs,showkeys]{revtex4-1}
\usepackage{amsmath,xspace}
\usepackage{amssymb}
\usepackage{epstopdf}
\usepackage{graphicx,color}

\newcommand{\degC}{\ensuremath{^{\circ}\text{C }}}
\newcommand{\sixrt}{\ensuremath{(6\sqrt{3}\!\times\!6\sqrt{3})_\text{SiC}\text{R}30^\circ}~}
\newcommand{\rt}{\ensuremath{6\sqrt{3}}}
\begin{document}
\title{Wide bandgap semiconductor from a hidden 2D incommensurate graphene phase}
	
\author{M.~Conrad}
\author{F.~Wang}
\author{M.S.~Nevius}
\affiliation{The Georgia Institute of Technology, Atlanta, Georgia 30332-0430, USA}
\author{K. Jinkins}
\affiliation{University of Wisconsin-Platteville, WI, 53818, USA}
\author{A. Celis }
\affiliation{Laboratoire de Physique des Solides, UniversitŽ Paris-Sud, CNRS, UMR 8502, F-91405 Orsay Cedex, France}
\author{M.N. Nair}
\affiliation{Synchrotron SOLEIL, L'Orme des Merisiers, Saint-Aubin, 91192 Gif sur Yvette, France}
\author{A. Taleb-Ibrahimi}
\affiliation{UR1 CNRS/Synchrotron SOLEIL, Saint-Aubin, 91192 Gif sur Yvette, France}
\author{A. Tejeda}
\affiliation{Laboratoire de Physique des Solides, UniversitŽ Paris-Sud, CNRS, UMR 8502, F-91405 Orsay Cedex, France}
\affiliation{Synchrotron SOLEIL, L'Orme des Merisiers, Saint-Aubin, 91192 Gif sur Yvette, France}
\author{Y. Garreau}
\author{A. Vlad} 
\author{A. Coati}
\affiliation{Synchrotron SOLEIL, L'Orme des Merisiers, Saint-Aubin, 91192 Gif sur Yvette, France}
\author{P.F.~Miceli}
\affiliation{Department of Physics and Astronomy, University of Missouri-Columbia, Columbia, MO 65211}
\author{E.H.~Conrad}\email{edward.conrad@physics.gatech.edu}
\affiliation{The Georgia Institute of Technology, Atlanta, Georgia 30332-0430, USA}


\begin{abstract}
  Producing a usable semiconducting form of graphene has plagued the development of graphene electronics for nearly two decades.  Now that new preparation methods have become available, graphene's intrinsic properties can be measured and the search for semiconducting graphene has begun to produce results.  This is the case of the first graphene ``buffer" layer grown on SiC(0001) presented in this work. We show, contrary to assumptions of the last forty years, that the buffer graphene layer is \textit{not} commensurate with SiC. The new modulated structure we've found resolves a long standing contradiction where \textit{ab initio} calculations expect a metallic buffer, while experimentally it is found to be a semiconductor. Model calculations using the new incommensurate structure show that the semiconducting $\pi$-band character of the buffer comes from partially hybridized graphene incommensurate boundaries surrounding unperturbed graphene islands.
\end{abstract}

\maketitle

Producing a technologically relevant semiconducting form of graphene has been a critical stumbling block towards graphene electronics that ultimately led research to shift to other less favorable 2D materials.  
The first graphene layer grown on the SiC(0001) surface, known as the ``buffer" layer, was considered to be an early candidate for graphene electronics because it was thought to be functionalized by bonding to the SiC surface.  
Angle resolved photoemission spectroscopy (ARPES) experiments on UHV grown samples found no graphene $\pi$-bands above the SiC valence band maximum indicating that the buffer was a wide gap semiconductor.\cite{Emtsev_MSF_07,Emtsev_PRB_08}  
However, significant surface states in the gap made device fabrication problematic. Theoretically, the nature of a semiconducting buffer remains unclear. Early calculations were based on graphene artificially strained to commensurately fit on smaller, computationally less demanding unit cells rather than the full \sixrt cell observed in low energy electron diffraction (LEED) (subsequently referred to as the \rt).\cite{VanBommel_SS_75,Forbeaux_PRB_98} 
These calculations found the buffer to be an insulator with a metallic state from Si dangling bonds in the SiC interface layer.\cite{Varchon_PRL_07,Mattausch_PRL_07} 
Calculations on the \rt~commensurate graphene-SiC system challenged the insulating buffer picture.\cite{Kim_PRL_08}  While not explicitly stated, the band structure presented by Kim et al.,\cite{Kim_PRL_08} showed that a modified network of $\pi$-bonds produced several bands dispersing through the Fermi level, i.e, the buffer layer was metallic.
In other words, graphene functionalized by the \rt ~bulk terminated reconstructed SiC surface was not sufficient to open a bandgap.

Interest in buffer graphene was recently renewed when it was discovered that buffer was a true semiconductor.\cite{Nevius_PRL_15} This form of the buffer can only be produced in a very narrow growth temperature range. ARPES measurements revealed that a gap opened in the buffer graphene $\pi$-bands with a promising effective mass and no observable surface states associated with previous buffer films.
It is now clear that no theoretical works, past or present, predict the observed band structure.  These facts, coupled with the large and computationally difficult \rt~unit cell, means that detailed structural information of the buffer-SiC interface is essential to understand the buffer's electronic properties.

In this work, we resolve the contradiction between the theoretical and experimental buffer band structure.  We do this by making the first high resolution surface x-ray diffraction (SXRD) measurements of the graphene-SiC(0001) structure and  coupling the results to a simple theoretical model to calculate the buffer graphene band structure.  The sample's coherent domain size is nearly double that of previous works\cite{Charrier_JApPhys_02,Hass_APL_06} allowing us to perform precise measurements of the in-plane surface structure. We demonstrate, contrary to the assumptions of the last four decades, that both the graphene buffer lattice \textit{and} the so-called \rt~``reconstruction" diffraction peaks are incommensurate with bulk SiC.  This observation leads to a new way to view the buffer graphene/SiC interface.  Rather than the conventional view of a reconstructed interface, we show that the correct picture is an incommensurate graphene lattice engaging in a mutual structural modulation with the \textit{bulk} SiC, where the modulation is composed of reciprocal lattice vectors of graphene and SiC.   The fundamental period of the modulation is found to be $\lambda=6(1+\delta)a_\text{SiC}$ where $\delta=0.037(2)$.  

We show that the distortion leads to a periodic hexagonal density gradient at the SiC-graphene interface that alters the graphene-SiC bonding symmetry.   Specifically, the incommensurate structure consists of weakly noninteracting graphene islands connected to a hexagonal network of incommensurate domain wall-like regions consisting of bonded and un-bonded graphene atoms similar to those seen by scanning tunneling microscopy (STM).\cite{Riedl_PRB_07}  Tight binding calculations based on this wall-like structure ultimately lead to a bandgap in the buffer's $\pi$-bands that is consistent with experimentally measured bands. 

Incommensurate crystals (IC) have well ordered periodic distortions that cannot be related by integer multiples to their underlaying lattice. 
X-ray diffraction has been used to study IC structures for over forty years\cite{vanLanduyt_PSSA_74} and remain the ideal technique to study IC systems, becuase unlike real space probes like STM, x-ray diffraction can precisely measure both absolute and relative small deviations from commensurate lattices.

\begin{figure}
	\centering
	\includegraphics[angle=0,width=8.0cm,clip]{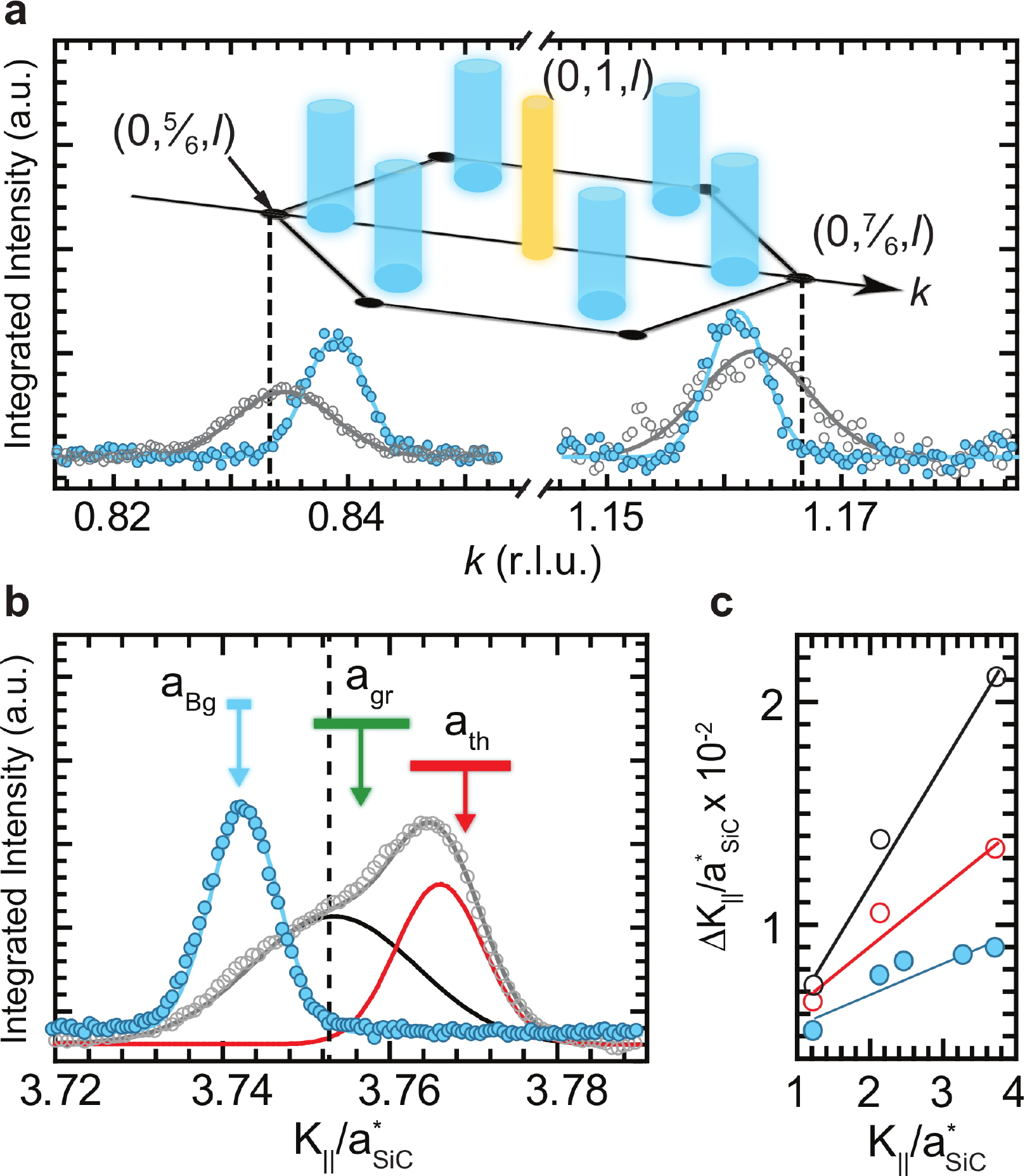}
	\caption{Diffraction results from the incommensurate graphene-SiC(0001) system.  (a) SXRD radial $k$ scans, $(0,k,0.1)$, around the SiC $(0,1,l)$ rod (see schematic in the insert). The background-subtracted intensity is instrument corrected\cite{Vlieg_JAC_97}. 
		Data is for the $\text{BG}_o$ (blue) and MG (grey) films.   Dashed lines mark the positions of the commensurate $5/6^\text{th}$ and $7/6^\text{th}$ diffractions rods (black circles in insert). The $\text{BG}_o$ satellite rods are contracted relative to the commensurate positions towards the $(0,1,l)$ rod. 
		(b) Radial scan through the nominal graphene $(0,3,0.1)_G$ rod for the $\text{BG}_o$ (blue $\circ$) and MG (grey $\circ$) films. Dashed line marks the expected position for a commensurate \rt~graphene film. Blue arrow shows the calculated $(0,3,l)_G$ position from Eq.~\ref{E:q}. The monolayer film has a contribution from the MG (red line) and the $\text{BG}_\text{ML}$ rods (black line). The green (red) arrow marks the position for graphite (theoretical graphene). The arrows' horizontal bar represent their known uncertainties. (c) Radial width of graphene rods as a function of $K_\parallel$  for $\text{BG}_o$ (blue $\circ$), MG (red $\circ$), and $\text{BG}_\text{ML}$ (grey $\circ$). }
	\label{F:Sat}
\end{figure}

To demonstrate how the buffer graphene (referred to as $\text{BG}_o$) lattice constant and electronic structure are a direct result of the incommensurate (IC) structure, we first quantify the IC distortion. In the traditional buffer layer picture, the commensurate \rt~ reconstruction gives rise to $6^\text{th}$ order diffraction rods at the $(0,\frac{5}{6},l)$ and $(0,\frac{7}{6},l)$ positions around the SiC $(01l)$ rod [see the insert in Fig.~\ref{F:Sat}(a)]. Experimentally, however, we find the satellite rods symmetrically shifted away from the commensurate positions towards the bulk $(01l)$ rod [see Fig.~\ref{F:Sat}(a)]. This is the classical result for an incommensurate system.\cite{vanLanduyt_PSSA_74}  

The contracted satellite peaks are a direct result of the commensurate in-plane unit cell positions, ${\bf R}$, being modulated by a function $\boldsymbol{\eta}({\bf R},{\bf q})$. The new modulated positions, ${\bf r}$, are incommensurate with ${\bf R}$ and given by,\cite{Overhauser_PRB_71, CUMMINS_PReport_90,Mato_JoPC1986} 
\begin{equation}
{\bf r}={\bf R}+\sum_{j=1}^d\boldsymbol{\eta}_{j}\sin{({\bf q}_j\cdot{\bf R}+\phi_j)}.
\label{E:R-mod}
\end{equation}
Here we have Fourier expanded $\boldsymbol{\eta}({\bf R},{\bf q})$ using a set of amplitudes, $\lbrace\boldsymbol{\eta}\rbrace$ (size $d$), corresponding to the set of modulation wavevectors $\lbrace{\bf q}\rbrace$.  This distortion produces satellite rods in reciprocal space at positions ${\bf K}$ given by;\cite{vanLanduyt_PSSA_74} 
\begin{equation}
{\bf K}={\bf G}+m\tilde{{\bf q}},\\
\label{E:satellite}  
\end{equation}
where $m$ is an integer, $\tilde{{\bf q}}$'s are any linear combination of the incommensurate wavevectors $\lbrace{\bf q}\rbrace$, and  ${\bf G}(h,k)$ is a reciprocal lattice vectors of the periodic lattice [see Methods for notation]. Experimentally the incommensurate rods along $k$ in Fig.~\ref{F:Sat}(a) are $\pm\tilde{{\bf q}}\!\equiv\!\pm{\bf q}_1\!=\!{\bf K}-{\bf G}^\text{(SiC)}_{0,1}$ whose magnitude is $|q_1|\!=\!a^*_\text{SiC}/6(1+\delta)$, where $\delta\!=\!0.037(2)$.

What is unique about the 2D buffer system is that the buffer and the SiC layer between the buffer and the bulk are mutually modulated.  We know this because the spacing between the buffer graphene ${\bf G}^\text{(g)}_{1,1}$ rod and the SiC ${\bf G}^\text{(SiC)}_{2,0}$ rod is also ${\bf q}_1$ (i.e., ${\bf q}_1\!=\!{\bf G}^\text{(SiC)}_{2,0}\!-\!{\bf G}^\text{(g)}_{1,1}$).  This coincidence occurs when the buffer distortion, ${\boldsymbol \eta}^\text{(g)}$, can be written as a Fourier expansion in terms of the SiC reciprocal lattice vectors ${\bf G}^\text{(SiC)}$, while the SiC interface distortion, ${\boldsymbol \eta}^\text{(SiC)}$, can be Fourier expanded in terms of the graphene reciprocal lattice vectors ${\bf G}^\text{(g)}$.  This leads to the condition that the satellite rods are given by:
\begin{equation}
\lbrace{\bf q}\rbrace = \pm ({\bf G}_i^\text{(SiC)}-{\bf G}_j^\text{(g)}).
\label{E:q}
\end{equation}

The mutual modulation provides two new and crucial insights into the physics of the graphene-SiC interaction. First, it accurately describes the measured graphene lattice constant, $a_\text{Bg}$. In the IC system we present, the correct buffer lattice constant can be determined from the measured ${\bf q}_1$ using Eq.~\ref{E:q} (${\bf G}_{1,1}^\text{(g)}\!=\!4\pi/a_\text{Bg}\!=\!{\bf G}_{2,0}^\text{(SiC)}\!-\!{\bf q}_1$).  Figure \ref{F:Sat}(b) compares the calculated $a_\text{Bg}$ with the measured graphene buffer $(0,3,0.1)_G$ rod.  The agreement is exceptional.  Note that the IC buffer graphene $(0,3,l)_G$ rod is shifted to lower $K_\parallel$ (larger lattice constant) compared to the expected position based on the commensurate \rt~cell (vertical dashed line). We point out that the buffer's lattice constant was previously measured to be incommensurate with SiC(0001) by Schumann et al.\cite{Schumann_PRB_14}  However, to reconcile their measured IC lattice constant with a commensurate \rt~cell Schumann et al.~attempted (incorrectly) to use a vertically buckled graphene sheet locked into the \rt~reconstruction.\cite{SupMat}  This model is completely inconsistent with the incommensurate satellite peaks we observe, which they did not measure.

The second insight of the mutual modulation is that it implies a significant interlayer bonding between the graphene and the SiC interface. The interlayer bonding must be sufficiently strong for the SiC interface to alter the graphene in-plane bonds, or conversely that the graphene in-plane bonds can structurally alter the SiC interface.
As we now show, the interaction and modulated geometry explain why the buffer is semiconducting instead of metallic.

\begin{figure}
	\centering
	\includegraphics[angle=0,width=8.0cm,clip]{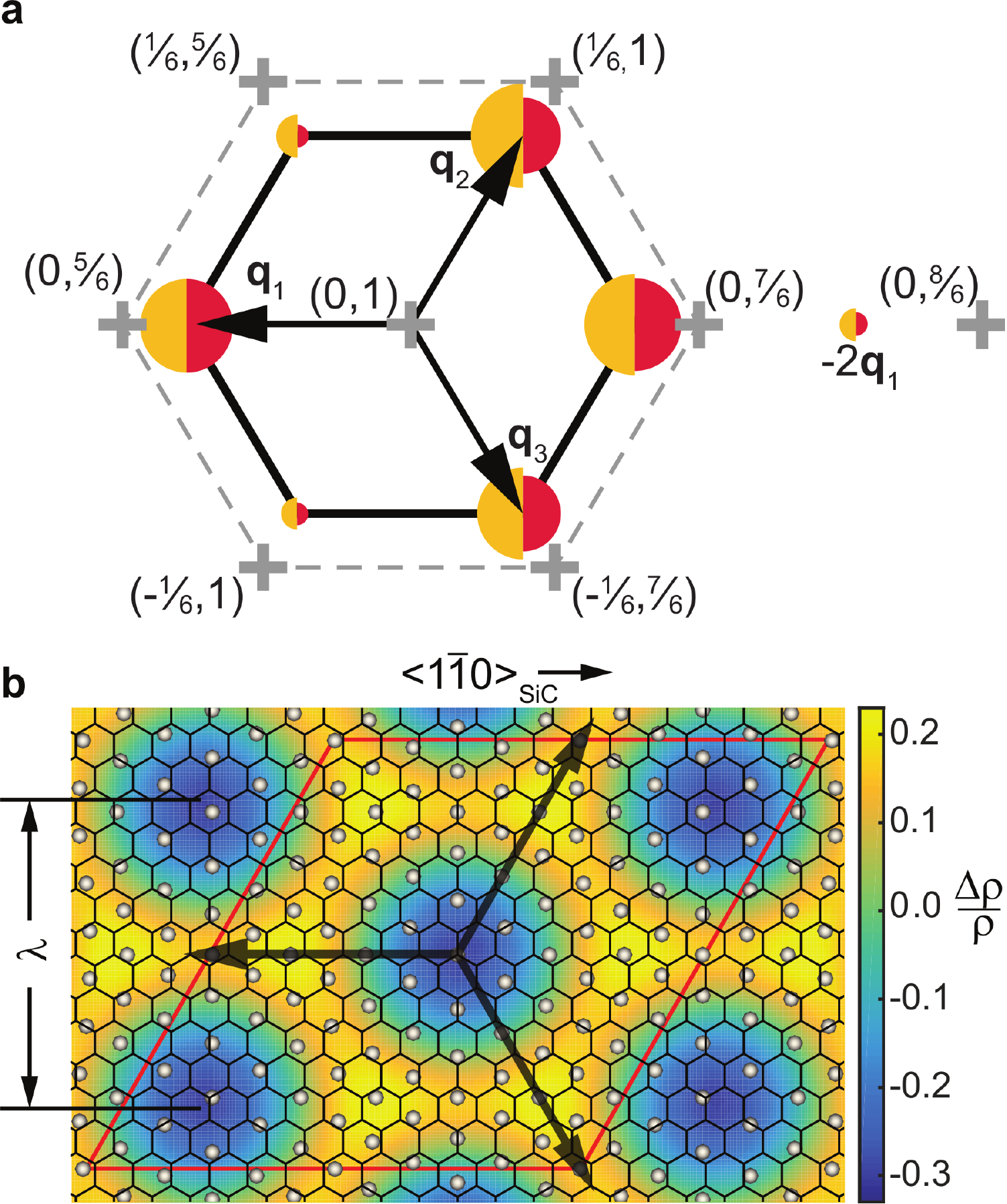}
	\caption{SXRD derived structure of the buffer-SiC interface.  (a) The instrument corrected\cite{Vlieg_JAC_97} integrated intensity of the satellite rods around the $(01l)$ rod. Crosses mark the commensurate $6^\text{th}$ order rods. The arrows show the three IC wavevectors.  The gold circle's area are proportional to the measured intensity of the satellite rods. The red circle's area are proportional to the fit intensity described in the text for $\eta^\text{(SiC)}\!=\!0.11a_\text{SiC}$. (b) Relative density $\Delta\rho(x,y)/\rho$ map of the incommensurate SiC interface using the measured ${\bf q}$'s and ${\boldsymbol \eta}^\text{(SiC)}$'s. The grey circles and hexagonal mesh overlay represents interface Si and graphene, respectively. The commensurate \rt~unit cell is marked in red.  Black arrows show the three IC wavevectors. }
	\label{BG2D}
\end{figure}

Calculating the scattered intensity from the IC lattice in Eq.~\ref{E:R-mod} is straightforward.\cite{SupMat}  However, because the satellite intensities are expected to decay rapidly with $|q|$,\cite{Overhauser_PRB_71} a meaningful comparison of the experimental data requires that we restrict the size of the set $\lbrace{\bf q}\rbrace$ to $d=3$. 
This choice is the minimum number of wavevectors necessary to reproduce the symmetry of the satellite intensities. The three ${\bf q}$'s are along equivlent graphene principle axes as shown in Fig.~\ref{BG2D}(a).  

By further assuming that $\lbrace\boldsymbol{\eta}\rbrace$ is isotropic and parallel to $\lbrace{\bf q}\rbrace$ we can estimate the modulation amplitude for both the SiC and buffer IC layers.  Figure \ref{BG2D}(a) shows the comparison of the measured to the modeled satellite intensities around the SiC $(0,1,l)$ rod. The modulation amplitude of the SiC interface layer is $\eta^{\text{(SiC)}}/a_\text{SiC}\!=\!0.11(4)$. A similar estimate for the $BG_o$ layer, using the intensities of the $(01)_G$ rod and its satellites, gives a much smaller, but non-zero buffer graphene in-plane modulation; $\eta^\text{(G)}/a_\text{G}\!\lesssim\!1\%$. 

The incommensurate distortion can be visualized by plotting the relative density $\Delta\rho/\rho$ of the SiC interface layer. 
The color map in Fig.~\ref{BG2D}(b) shows that the SiC interface consists of a super-hexagonal network with a period $\lambda\!=\!6(1+\delta)a_\text{SiC}$.  The network boundaries have a higher density than bulk terminated SiC.  Note that while the density modulation is periodic, the positions of the atoms given by Eq.~\ref{E:R-mod} in both the SiC interface and the buffer graphene are not periodic. The network is very similar to STM images of the buffer layer.\cite{Riedl_PRB_07,Goler2013}

The exact structure of the interface and the driving force for the IC transition remains to be determined.  It is unlikely that a simple sine wave distortion used to fit the data is a complete picture. An energetically more favorable structure would be a network of high density IC domain walls.  The data does not address what drives the distortion. However, recent work by Emery et al.\cite{Emery_SW_13} on UHV grown multilayer graphene films, may provide a clue.  They show that the interface layer below the buffer has a lower silicon and a higher carbon concentration than bulk SiC. Silicon vacancies could give rise to different carbon bonding geometries (e.g. C-vacancy or C-C bonds with $\text{sp}^2$ character) that could produce strains sufficient to drive the IC modulation in the SiC bilayer. 

\begin{figure}
	\centering
	\includegraphics[angle=0,width=14.0cm,clip]{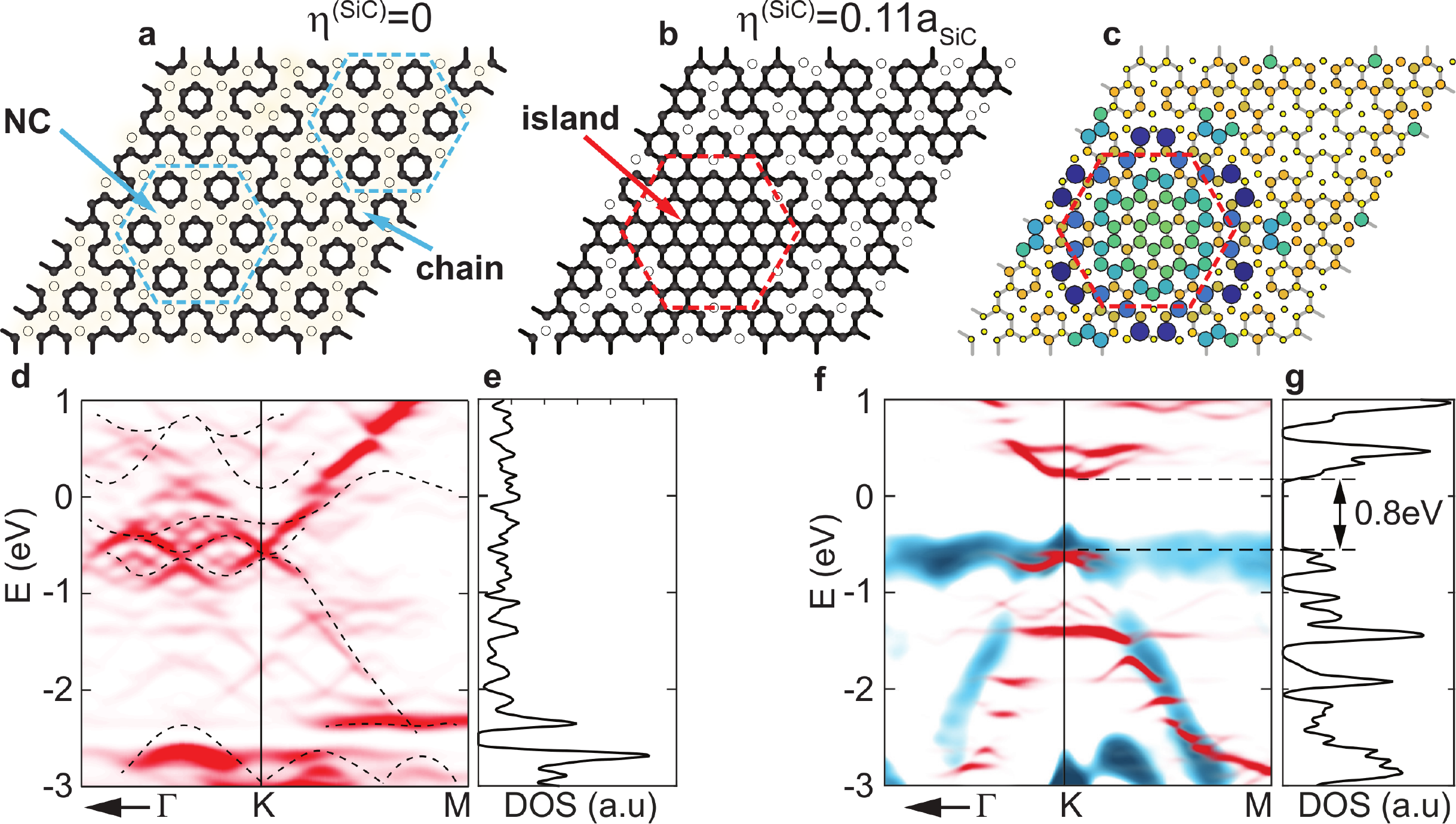}
	\caption{Comparison of the theoretical and the incommensurate graphene band structure with experimental ARPES data.
		(a) The commensurate \rt~buffer structure derived from \textit{ab initio} calculations in Ref.~[\citenum{Kim_PRL_08}]. Black circles are carbon unbonded to the SiC. Gold circles are carbon bonded to Si in the interface layer below. The NC regions (blue hexagons) and the carbon chains are marked.  (b) A model structure based on modulated SiC layer using the experimental value, $\eta^{(SiC)}=0.11a_\text{SiC}$ (same color scheme as (a)).  Red dashed hexagon marks the boundary of an isolated graphene island. (c) The calculated charge density\cite{Vanevic_PRB_09} (arbitrary units) at $E\!=\!-0.6$~eV for the structure in (b).  
		(d) TB  bands (red) mapped onto the graphene BZ\cite{Deretzis_EPL_2014} from the commensurate structure in (a). The low energy bands from the \textit{ab initio} commensurate structure are overlaid (black dashed line).  
		(e) DOS for the TB bands in (d).
		(f) TB calculated bands (red) from the modulated structure in (b).  The negative $2^\text{nd}$ derivative of the experimental ARPES bands (blue) are overlaid. The $\pi$-bands from a 2\% monolayer have been subtracted from the experimental bands. 
		(g) DOS for the TB bands in (f). The direct 0.8eV bandgap is marked.}
	\label{Theory}
\end{figure}

Regardless of the exact structure, the discovery of an incommensurately modulated interface allows us to revisit and explain the origin of the buffer layer's electronic structure. Previous \textit{ab initio} calculations of the buffer layer assumed a commensurate bulk-terminated SiC interface with a \rt~ unit cell.\cite{Kim_PRL_08} They found that 79\% of the interface Si atoms bind to 25\% of the $\text{BG}_o$ graphene C atoms [see Fig. \ref{Theory}(a) and Supplemental Material\cite{SupMat}]. The bonding pattern is divided into two types: a nearly commensurate (NC) region and partially bonded carbon chains.  In the NC region most of the carbon is bonded to silicon surface atoms, the remaining carbon in this region forms isolated benzene-like rings [see Fig.~\ref{Theory}(a)]. The chains are an incomplete hexagonal network of carbon not bonded to the SiC. 
This network creates boundaries around the NC regions that are responsible for the bands near $E_F$ [see Fig. \ref{Theory}(d)].\cite{Kim_PRL_08}

To study the electronic structure of the incommensurate system, we developed a tight binding model based on \textit{ab initio} calculations\cite{Kim_PRL_08}. Tight binding methods are often employed to study IC systems because of the large number of atoms involved.  These IC calculations involve a ``unit cell" arising from either a truncated lattice or by using a nearly commensurate lattice.\cite{Bak1982,Voit2000,SOKOLOFF1985,Ahmadieh_PRB_73,Laissardiere2010}. Although the Brillouin zone collapses in IC systems, delocalized dispersive states are still predicted for certain interaction configurations\cite{SOKOLOFF1985,Sokoloff1981}. Indeed APRES measurements of IC systems still show ``bands" .\cite{Voit2000} 

The predicted metallic band structure can be recreated in a simplified tight binding (TB) calculation of the buffer graphene $\pi$-bonds. In this model, a Si interface atom bonds to the nearest C atom in the buffer graphene if the in-plane C-Si distance is within a maximum radius, $R_\text{max}$ ($R_\text{max}$ is determined from the \textit{ab initio} calculations). The buffer $\pi$-orbitals of the bonded carbon atoms are assigned an onsite potential consistent with the \textit{ab initio} calculations and tight binding parameter estimates.\cite{SupMat} The excellent agreement between the \textit{ab initio} and TB band structure is shown in Fig. \ref{Theory}(d). While both calculations are in agreement, they clearly do not predict the ARPES experimental semiconducting bands plotted in Fig.~\ref{Theory}(f).

Within this TB ansatz (with a similar $R_\text{max}$ and on-site potential), we have explored the effect of a incommensurately modulated SiC lattice on the buffer's band structure. In order to compare our \rt~calculations with measured ARPES bands, we begin with a commensurate \rt~unit cell.
Significant changes occur in the C-Si bonding configuration when the bulk terminated surface is modulated according to Eq.~\ref{E:R-mod}.\cite{SupMat} Figure \ref{Theory}(b) shows the bonding structure using the experimentally measured $\eta^\text{(SiC)}$. The modulation decreases the number of Si bonds to the buffer graphene layer by nearly 40\% compared to the commensurate case in Fig.~\ref{Theory}(a). Half of the NC regions in the commensurate structure converts into large regions of unperturbed graphene ``islands'' corresponding to half the low density regions in Fig.~\ref{BG2D}(b).  The graphene between the islands, aligned with the high density boundaries in Fig. \ref{BG2D}(b), have a much higher number of bonds to the interface Si as might be expected. The interface density modulation acts as a domain wall in the buffer graphene layer that breaks the bonding symmetry and opens a band gap. The formation of islands and the opening of a band gap occur for both larger ``incommensurate" unit cells and  over a large range of  $\eta$'s ($0.05\!<\!\eta^\text{(SiC)}/a_\text{SiC}\!<\!0.36$) that includes the experimentally determined value of $\eta^\text{(SiC)}$. The gap increases as a function of $\eta^\text{(SiC)}$ becoming nearly constant for $\eta^\text{(SiC)}/a_\text{SiC}\gtrsim\!0.1$.\cite{SupMat}

The calculated semiconducting bands [see Fig. \ref{Theory}(f)] look remarkably similar to the measured ARPES bands for $\eta^\text{(SiC)}\!=\!0.11a_\text{SiC}$ (the experimental value). The predicted gap is $~0.8$ eV as shown in density of states (DOS) in Fig. \ref{Theory}(g). The charge density from the three highest occupied bands in Fig.~\ref{Theory}(f) show weak localization at the edges of the island [see Fig. \ref{Theory}(c)] and give rise to a charge density remarkably similar to STM measurements.\cite{Riedl_PRB_07} 

When a monolayer of graphene (MG) forms above the buffer, there are changes in the buffer's structure and electronic properties. We refer to the buffer with MG on top as $\text{BG}_\text{ML}$ to distinguish it from the bare buffer layer $\text{BG}_o$. While it is known that the buffer layer contracts when the monolayer forms [see Fig.~\ref{F:Sat}(b)],\cite{Schumann_PRB_14} the reason for the change remains conjecture. SXRD shows that the buffer's strain is due to a change in the modulation wavelength.  Once the MG forms, the satellite rod positions and the $\text{BG}_\text{ML}$ lattice constant contract to a film nearly commensurate with the bulk SiC (i.e, $\delta\!\sim\!0.02$)  [see Fig.~\ref{F:Sat}(a)].  The MG lattice constant contracts relative to the $\text{BG}_\text{ML}$ making the MG incommensurate with both the $\text{BG}_\text{ML}$ and the SiC. The lattice constants for buffer and MG systems are summarized in Table  \ref{tab:Lattice_a}.  

\begin{table}
	\caption{\label{tab:Lattice_a} Comparisons of graphene lattice constants, their relative strain ($\Delta a$) compared to theoretical graphene, RMS strain $\epsilon_{\text{rms}}$, and long range order}
	\begin{tabular}{llccccc}
		Graphene & Lattice & $\Delta a$ & $\epsilon_{\text{rms}}$ & Order \\
		Form  & constant (\AA) & (\%) &(\%) & (nm)\\
		\hline 
		Theoretical MG 		& $2.453(4)$\textsuperscript{\emph{4}}				&- 			&- 		&-\\
		Graphite            		& $2.460(2)$\textsuperscript{\emph{5}}				& +0.28 		&- 		&-\\
		$\text{BG}_o$ 		& $2.469(3)$\textsuperscript{\emph{1},}\textsuperscript{\emph{3}} 	& +0.70  		& 0.2 	&60\\
		$\text{BG}_\text{ML}$ 	& $2.462(3)$\textsuperscript{\emph{1},}\textsuperscript{\emph{3}} 	& +0.40 		& 0.6  	& 43\\
		MG 					& $2.455(3)$\textsuperscript{\emph{1},}\textsuperscript{\emph{3}} 	&+0.10 		& 0.3 	& 43\\
		C-Face multilayer      & $2.452(3)$\textsuperscript{\emph{2}}   		&  -0.04  &  -   & 300   \\
		
	\end{tabular}
	\begin{flushleft}
		\textsuperscript{\emph{1}} This work\\
		\textsuperscript{\emph{2}}  From Ref. [\citenum{Hass_PRL_08}].\\
		\textsuperscript{\emph{3}} Similar values were measured by \citet{Schumann_PRB_14}.\\
		\textsuperscript{\emph{4}} From Refs. [\citenum{Weinert_PRB_82,Sahin_PRB2009,Wang_JPSoJ_10}].\\
		\textsuperscript{\emph{5}} From Refs. [\citenum{Nelson_PPhysSoc_45,Baskin_PR_55,Bosak_PRB_07,Ahmadieh_PRB_73}].
	\end{flushleft}
\end{table}

Note that the MG lattice constant is within the range of values reported for theoretically isolated graphene,
 2.453(4)\AA.\cite{Weinert_PRB_82,Sahin_PRB2009,Wang_JPSoJ_10} This is consistent with models of weakly coupled layered materials (including graphite) where stronger interlayer interaction cause larger in-plane expansion.\cite{Barrera_JPCM_05}  The MG is contracted relative to graphite both because it interacts with a single layer and because the incommensuration between the two layers reduces the number of inter-layer bonds compared to Bernal stacking. This effect is analogous to the in-plane contraction observed in non-Bernal stacked graphene layers on C-face SiC [see Table \ref{tab:Lattice_a}].\cite{Hass_PRL_08}  

The measured lattice constants show that graphene's intrinsic strain has been historically misinterpreted using Raman 2D peak positions. It has been clearly demonstrated that the Raman 2D peak of tensile strained graphene red shifts to lower wave numbers.\cite{Zabel_NL_12}  Since isolated graphene must be compressively strained relative to graphite [see Table \ref{tab:Lattice_a}], the 2D peak of true free standing graphene must be blue shifted relative to graphite as is the case for the MG 2D peak.\cite{Ni_PRB_08,Rohrl_APL_08,Speck_APL_11,Fromm_NJP_13}  
The problem is that exfoliated graphene has its 2D peak shifted in the \emph{wrong} direction (it is red shifted compared to graphite),\cite{Ni_PRB_08,Rohrl_APL_08,Speck_APL_11,Fromm_NJP_13} which contradicts its historical reference as ``free standing" graphene. Clearly the position of the 2D Raman peak in ``free standing" graphene is due to some other cause that has yet to be explained.

There are two other significant changes when the MG forms.  First, the system becomes more disordered (30\% decrease in long range order) evidenced by the broader satellite rods in Fig.~\ref{F:Sat}(b).   The $\text{BG}_\text{ML}$ also develops a large RMS strain, $\epsilon_\text{rms}$.  RMS strain presents itself as a $K$-dependent broadening ($\Delta K\!\approx\!\epsilon_\text{rms} K$).  As Fig.~\ref{F:Sat}(c) shows, the $\text{BG}_\text{ML}$ has the largest slope (largest $\epsilon_\text{rms}$) in a plot of $\Delta K$ vs $K$ [$\epsilon_\text{rms}$ data is summarized in Table \ref{tab:Lattice_a}]. The MG has a smaller RMS strain, presumably due to the weak coupling to the $\text{BG}_\text{ML}$ layer that allows strain relaxation. 

\begin{figure}
	\centering
	\includegraphics[angle=0,width=16cm,clip]{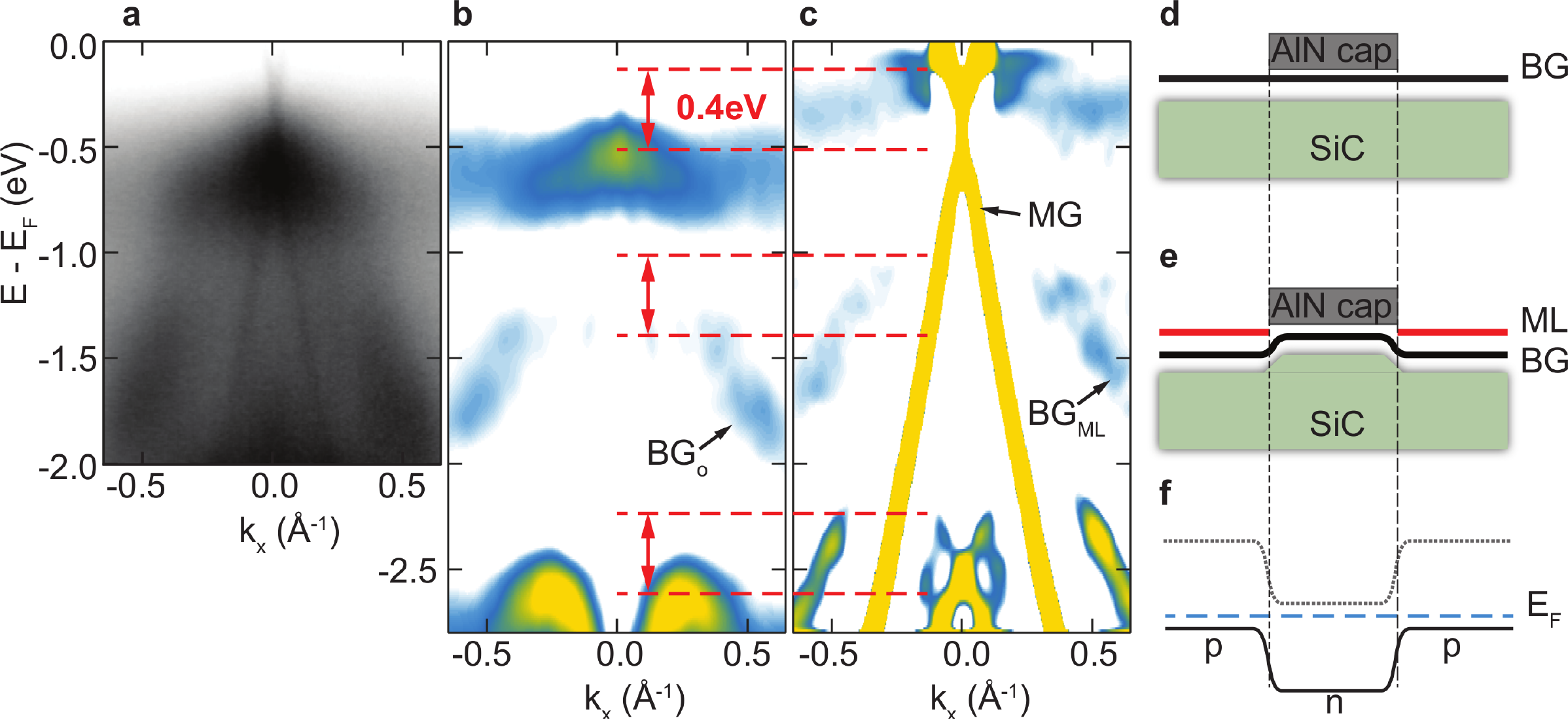}
	\caption{The effect of ML graphene growth on the buffer band structure. (a) ARPES bands at the $\text{BG}_o$ layer K point ($k_x$ is perpendicular to $\Gamma K$, $h\nu\!=\!70$~eV).  A Dirac cone from a 2\% ML graphene layer is also visible.  (b) A negative $2^{nd}$ derivative filter of the $\text{BG}_o$ bands in (a).  (c) A similar $2^{nd}$ derivative filter for a MG film. Red dashed lines mark the approximate 0.4~eV shift of the buffer bands. (d) Schematic of a negative AlN capping layer to locally prevent graphene growth.  (e) Schematic of a pnp junction made by monolayer-buffer-monolayer junction where the buffer layer is a continuous film. (f) Schematic of the spatially varying bands from the structure in (e).}
	\label{f:ARPES}
\end{figure}

Finally we address the buffer layer's stability.  It is assumed that the strong buffer-SiC interaction meant that the buffer's band structure did not change significantly once the MG formed.  Now that we have shown that there is a structural change in the buffer layer when the MG forms, it is prudent to revisit how or if the $\text{BG}_\text{ML}$ electronic structure is different from the $\text{BG}_o$ layer. Figure \ref{f:ARPES}(a) shows the ARPES  spectra from the $\text{BG}_o$ layer. The $\pi$-bands are broad ($\Delta k\!\sim\! 0.35~\text{\AA}^{-1}$) consistent with the IC wavevector $q\sim\!0.38~\text{\AA}^{-1}$. In order to compare the $\text{BG}_\text{ML}$ bands with the $\text{BG}_o$, we have plotted a $2^\text{nd}$ derivative spectra of the buffer and MG bands in Fig.~\ref{f:ARPES}(b) and (c). This compensates for both the $\Delta k$ broadening and the photoelectron attenuations through the MG. 

Figure \ref{f:ARPES}(c) shows that the semiconducting $\pi$-bands are still present with the MG above.  Although the $\text{BG}_\text{ML}$ bands intensity is weak, it is consistent with a complete buffer layer after correcting for attenuation. There is, however a change in the $\text{BG}_\text{ML}$ bands compared to the $\text{BG}_o$ bands.  The $\pi$-bands are pushed to lower binding energy by $\sim\!0.4$eV compared to the $\text{BG}_o$ bands and the band near $E_F$ appears to have less dispersion than the $\text{BG}_o$ case.  While there is a small energy gap between the $\text{BG}_\text{ML}$ layer bands and $E_F$, the experimental error could also support the $\text{BG}_\text{ML}$ layer being metallic.  We note that $\eta^\text{(SiC)}\!<0.05~a_{SiC}$ (the uncertainty is due to the increased disorder in the $\text{BG}_\text{ML}$).  The low value of $\eta^\text{(SiC)}$ is consistent with a buffer layer structure closer to the commensurate structure that would give rise to either a very small gap or metallic bands.

The band changes in the $\text{BG}_\text{ML}$ layer suggests a pnp junction device architecture by spatially controlling where MG is formed [see Fig.~\ref{f:ARPES}(d), (e), and (f)]. By first growing a continuous buffer layer, a capping layer (AlN or SiN) mask is locally deposited to inhibit further graphene growth.\cite{Roy_APL_10,Puybaret_JPhyD_15}   The sample is then heated to grow MG outside of the masked area.  This leaves a pnp junction as shown in Fig.~\ref{f:ARPES} where the capping layer remains as a gate insulator.

Our x-ray measurements shows that the conventional assumption of a graphene film on SiC(0001) being commensurate with  \sixrt reconstruction is wrong.  We find instead that first graphene ``buffer"  layer and a SiC interface layer have a mutual incommensurate density modulation. The modulation wavevector, $|q|\!=\!a^*_\text{SiC}/6(1+0.037)$ gives rise to diffraction rods near (but not at) the \rt~positions explaining the misinterpretation of low resolution LEED data.  The discovery answers a number of persistent and important questions about electronic and transport properties of both the buffer and the monolayer that grows above it. Using a simple model based on the experimental parameters from the incommensurate modulation, we show that the tight binding derived band structure is remarkably similar to the experimentally measured semiconducting bands.  Our finding explain why \textit{ab initio} calculations, based on a commensurate buffer-SiC system,\cite{Kim_PRL_08} predicts metallic $\pi$-bands instead of the experimentally determined semiconductor bands.\cite{Emtsev_MSF_07,Nevius_PRL_15}    

We also find that when a monolayer graphene layer grows on top of the buffer, it is incommensurate with the buffer.  This incommensuration gives insight into why the monolayer graphene has a mobility an order of magnitude smaller compared to C-face graphene.\cite{Berger06,Speck_MSF_10,Jobst_PRB_10}  We suggest that because the monolayer is not periodic with the buffer below, a network of quasi-random monolayer-buffer interlayer coupling sites develop. These sites will increase random scattering in the film and thus lower the monolayer mobility.  We also show that the buffer structure and corresponding band structure change when a monolayer grows above it. The band changes implies that a pnp junction can be created by locally growing a monolayer on an otherwise continuous buffer layer film. This effect suggests a new method for band gap engineering in graphene systems and potentially new architectures for graphene switching devices.

\section*{Author Contributions}
E.H.C. supervised the project, designed and carried out the experiment, analyzed data and wrote the manuscript. M.C prepared samples, performed SXRD and ARPES experiments, analyzed data, performed TB calculations, and helped write the manuscript. F.W. performed SXRD experiments and analyzed data.  M.S.N. prepared samples and edited the manuscript.  K.J. helped analyze SXRD data. A. Ce., M.N.N., A.T-I, and A.T. performed ARPES experiments.  Y.G., A.V. and A.Co helped perform SXRD experiments and critical reads of the manuscript.  P.F.M. performed SXRD experiments, developed critical analysis of the SXRD results and critical reviews of the manuscript.

\section*{Experimental}
The substrates used in these studies were n-doped CMP polished on-axis 4H-SiC(0001). The graphene was grown in a confinement controlled silicon sublimation furnace.\cite{WaltPNAS} In the CCS method, graphene growth is a function of temperature, time, and crucible geometry that sets the silicon vapor pressure. With the current crucible design, a single semiconducting buffer graphene layer grows at a temperature of 1400\degC for 30~min while a MLG film will grow above a buffer layer at 1560\degC in 20~min. An important note is that once the MLG grows, a new buffer layer forms below the MLG layer. As we demonstrate, the buffer graphene with and without a MLG layer above it will be structurally different. The $\text{BG}_o$ and MLG samples were characterized by Raman spectroscopy and LEED and give results similar to Fromm et al.\cite{Fromm_NJP_13} and Emtsev et al.\cite{Emtsev_PRB_08}. 

SXRD  measurements were conducted at room temperature under UHV at the SIXS beamline at SOLEIL Synchrotron using a photon energy of $h\nu\!=\!12.8$~keV. The angle of incidence was fixed at $0.1^\circ$ (near the critical angle) to optimize the scattered intensity. The momentum transfers, ${\bf K}\!=\!{\bf k}_f\!-\!{\bf k}_i$, are written in terms of the bulk hexagonal SiC lattice parameters: ${\bf K}\!=\!h{\bf a}^*_{\text{SiC}}\!+\!k{\bf b}^*_{\text{SiC}}\!+\!l{\bf c}^*_{\text{SiC}}$ where $a^*_{\text{SiC}}\!=\!b^*_{\text{SiC}}\!=\!2\pi/\left(a_{\text{SiC}}\sqrt{3}/2\right)\!=\!2.3556\AA^{-1}$ and $c^*_{\text{SiC}}\!=\!0.6233\AA^{-1}$. ${\bf K}=\left(h,k,l\right)$ represents a point in reciprocal space using SiC reciprocal lattice units (r.l.u.). Polarization and geometric corrections\cite{Vlieg_JAC_97} were performed to compare integrated intensities. Prior to X-ray exposure, the samples were heated to 500\degC in UHV to remove surface contaminants. 

ARPES measurements were made at the Cassiop\'ee beamline at the SOLEIL synchrotron. Measurements on both samples were performed at 90 K. The Fermi energy, $E_{F}$, was determined from the k-integrated intensity cutoff of the molybdenum sample holder to within 20 meV for each sample. $h\nu\!=\!70$~eV.

\section*{acknowledgement}
	This research was supported by the National Science Foundation under Grant No. DMR-1401193. P.F. Miceli also acknowledges support  from the NSF under grant No DGE-1069091.   Additional support came from the Partner University Fund from the French Embassy. The authors thank W. de Heer for use of the Keck Lab facilities, and A. Zangwill and M. Mourigal for many helpful discussions.  We also wish to thank M. Kindermann and S. Spitz for help in the TB calculations.  Finally, we with to thank the staff and technical support given for this project by the Synchrotron Soleil.

\section*{Supplemental Information}

\begin{itemize}
  \item Supplemental Material: Diffraction intensity derivation, tight binding model and the incommensurate buffer graphene lattice constant
\end{itemize}

\bibliography{library}
\end{document}